\documentclass[aps,prb,preprint,superscriptaddress]{revtex4-1}
\usepackage{natbib}
\usepackage{graphicx}
\usepackage{subfigure}
\usepackage{bm}
\begin{document}
\title{Hard particle spectra of galactic X-ray sources by relativistic magnetic reconnection in laser lab}
\author{K.F.F. Law$^{*}$}
\affiliation{Institute of Laser Engineering, Osaka University, 2-6 Yamadaoka, Suita, Osaka 565-0871, Japan}
\author{Y. Abe}
\affiliation{Institute of Laser Engineering, Osaka University, 2-6 Yamadaoka, Suita, Osaka 565-0871, Japan}
\author{A. Morace}
\affiliation{Institute of Laser Engineering, Osaka University, 2-6 Yamadaoka, Suita, Osaka 565-0871, Japan}
\author{Y. Arikawa}
\affiliation{Institute of Laser Engineering, Osaka University, 2-6 Yamadaoka, Suita, Osaka 565-0871, Japan}
\author{S. Sakata}
\affiliation{Institute of Laser Engineering, Osaka University, 2-6 Yamadaoka, Suita, Osaka 565-0871, Japan}
\author{S. Lee}
\affiliation{Institute of Laser Engineering, Osaka University, 2-6 Yamadaoka, Suita, Osaka 565-0871, Japan}
\author{K. Matsuo}
\affiliation{Institute of Laser Engineering, Osaka University, 2-6 Yamadaoka, Suita, Osaka 565-0871, Japan}
\author{H. Morita}
\affiliation{Institute of Laser Engineering, Osaka University, 2-6 Yamadaoka, Suita, Osaka 565-0871, Japan}
\author{Y. Ochiai}
\affiliation{Institute of Laser Engineering, Osaka University, 2-6 Yamadaoka, Suita, Osaka 565-0871, Japan}
\author{C. Liu}
\affiliation{Institute of Laser Engineering, Osaka University, 2-6 Yamadaoka, Suita, Osaka 565-0871, Japan}
\author{A. Yogo}
\affiliation{Institute of Laser Engineering, Osaka University, 2-6 Yamadaoka, Suita, Osaka 565-0871, Japan}
\affiliation{PRESTO, Japan Science and Technology Agency, 4-1-8 Honmachi, Kawaguchi, Saitama 332-0012, Japan}
\author{K. Okamoto}
\affiliation{Institute of Laser Engineering, Osaka University, 2-6 Yamadaoka, Suita, Osaka 565-0871, Japan}
\author{D. Golovin}
\affiliation{Institute of Laser Engineering, Osaka University, 2-6 Yamadaoka, Suita, Osaka 565-0871, Japan}
\author{M. Ehret}
\affiliation{Universit\'e de Bordeaux, CNRS, CEA, CELIA (Centre Lasers Intenses et Applications), UMR 5107, Talence, France}
\affiliation{Institut f\"ur Kernphysik, Technische Universit\"at Darmstadt, D-64289 Darmstadt, Germany}
\author{T. Ozaki}
\affiliation{National Institute for Fusion Science, National Institutes of Natural Sciences, 322-6 Oroshi-Cho, Toki, Gifu 509-5292, Japan}
\author{M. Nakai}
\affiliation{Institute of Laser Engineering, Osaka University, 2-6 Yamadaoka, Suita, Osaka 565-0871, Japan}
\author{Y. Sentoku}
\affiliation{Institute of Laser Engineering, Osaka University, 2-6 Yamadaoka, Suita, Osaka 565-0871, Japan}
\author{J.J. Santos}
\affiliation{Universit\'e de Bordeaux, CNRS, CEA, CELIA (Centre Lasers Intenses et Applications), UMR 5107, Talence, France}
\author{E. d'Humi\`eres}
\affiliation{Universit\'e de Bordeaux, CNRS, CEA, CELIA (Centre Lasers Intenses et Applications), UMR 5107, Talence, France}
\author{Ph. Korneev}
\affiliation{National Research Nuclear University MEPhI (Moscow Engineering Physics Institute), 31 Kashirskoe shosse, Moscow, 115409, Russian Federation}
\affiliation{P. N. Lebedev Physics Institute, Russian Academy of Sciences, 53 Leninskiy Prospekt, Moscow, 119991, Russian Federation}
\author{S. Fujioka$^{*}$}
\affiliation{Institute of Laser Engineering, Osaka University, 2-6 Yamadaoka, Suita, Osaka 565-0871, Japan}
\maketitle
\section{Abstract}
Magnetic reconnection is a process whereby magnetic field lines in different directions ''reconnect'' with each other, resulting in the rearrangement of magnetic field topology together with the conversion of magnetic field energy into the kinetic energy (K.E.) of energetic particles\cite{Yamada2010}. This process occurs in magnetized astronomical plasmas, such as those in the solar corona, Earth's magnetosphere, and active galactic nuclei, and accounts for various phenomena, such as solar flares\cite{Sweet1958}, energetic particle acceleration\cite{Hughes1995}, and powering of photon emission\cite{Di1998}. In the present study, we report the experimental demonstration of magnetic reconnection under relativistic electron magnetization situation, along with the observation of power-law distributed outflow in both electron and proton energy spectra. Through irradiation of an intense laser on a ''micro-coil'', relativistically magnetized plasma was produced and magnetic reconnection was performed with maximum magnetic field 3~kT. In the downstream outflow direction, the non-thermal component is observed in the high-energy part of both electron and proton spectra, with a significantly harder power-law slope of the electron spectrum ($p=1.535\pm0.015$) that is similar to the electron injection model\cite{Zdziarski2014} proposed to explain a hard emission tail of Cygnus X-1, a galactic X-ray source with the same order of magnetization. The obtained result showed experimentally that the magnetization condition in the emitting region of a galactic X-ray source is sufficient to build a hard electron population through magnetic reconnection.

\section{Main}
\begin{figure}[htbp]
    \centering\includegraphics[width=\linewidth]{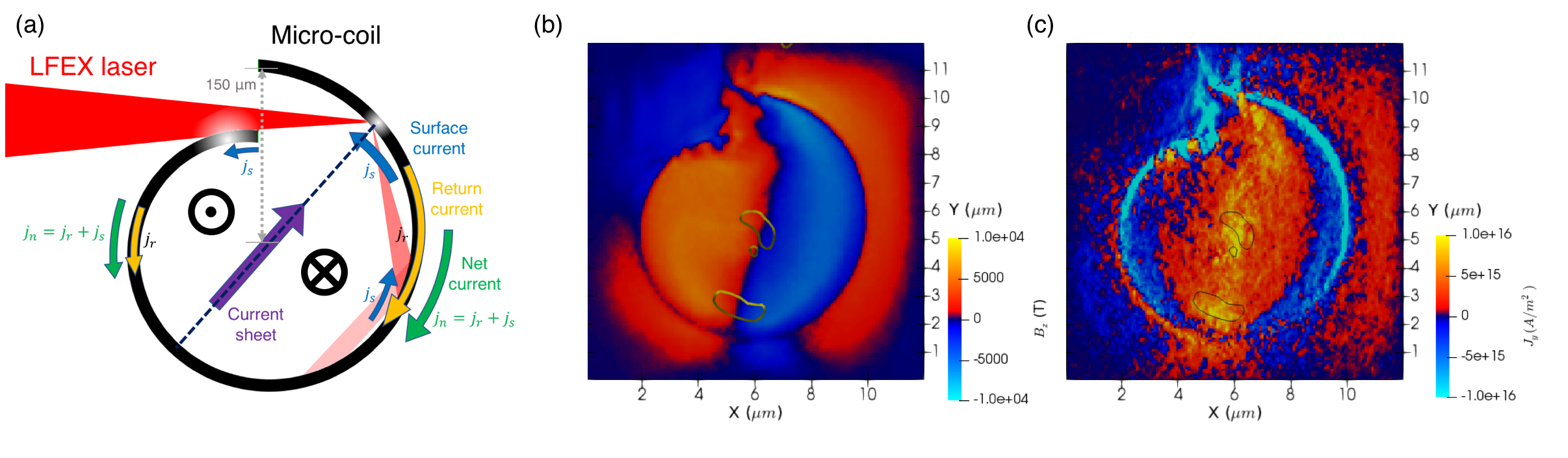}
    \caption{(a) Incidence laser intentionally focused on the inner surface of the micro-coil, such that a portion of laser is blocked and irradiated on the other end of the micro-coil. Bi-directional current (indicated by green arrows) is then generated. (b) The 3D-PIC simulation results (for 1/30 downscaled target size) showed that two directions of the magnetic field $B_z$ (component perpendicular to the cross section) formed inside the single micro-coil, reconnecting within the boundary between the two regions of anti-parallel $B_z$. The contour of electron dissipation measure\cite{Zenitani2011} $D_e$ is plotted by yellow lines, indicates magnetic reconnection sites. (c) Current sheet $J_y$ in same simulation, plotted together with the contour of $D_e$. Value of $D_e$ is normalized by $v_{A0}B_0j_0$, $v_{A0}$, $B_0$ and $j_0$ be Alfv\'en velocity, magnetic field and current density in simulation respectively, as previous study\cite{Zenitani2011}.}
    \label{fig:fig1}
\end{figure}

\begin{figure}[htbp]
\centering\includegraphics[width=\linewidth]{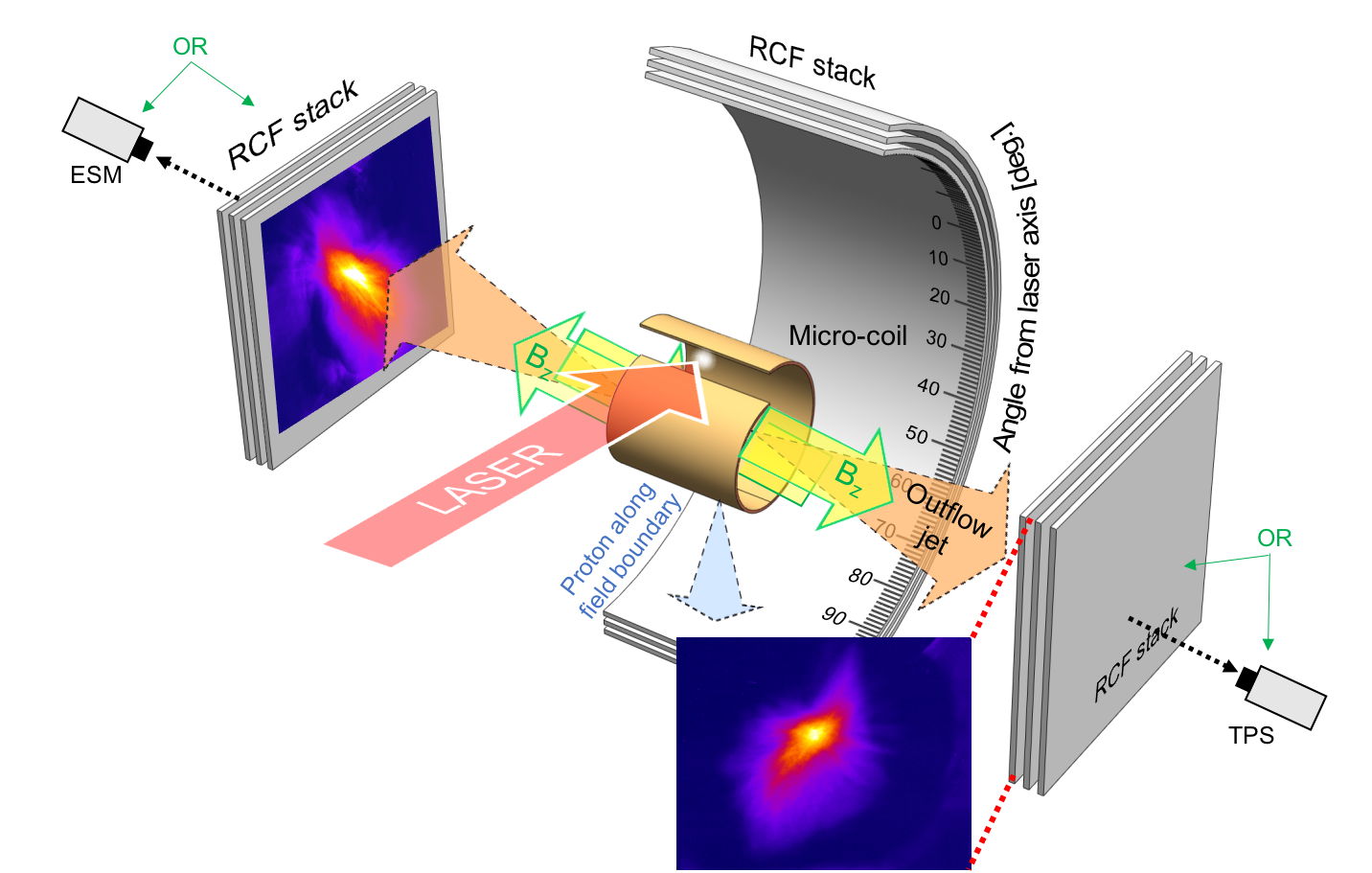}
    \caption{Observation of a pair of reconnection jets accelerated from the micro-coil. A pair of RCF stacks was used to detect the angular distribution of the protons in the jets. The two proton patterns, which appeared to be symmetric, are obtained by the same laser shot. Protons are accelerated and guided along the anti-parallel magnetic field boundary and in the direction of current sheet.}
    \label{fig:fig2}
\end{figure}

\begin{figure}[htbp]
  \centering\includegraphics[width=\linewidth]{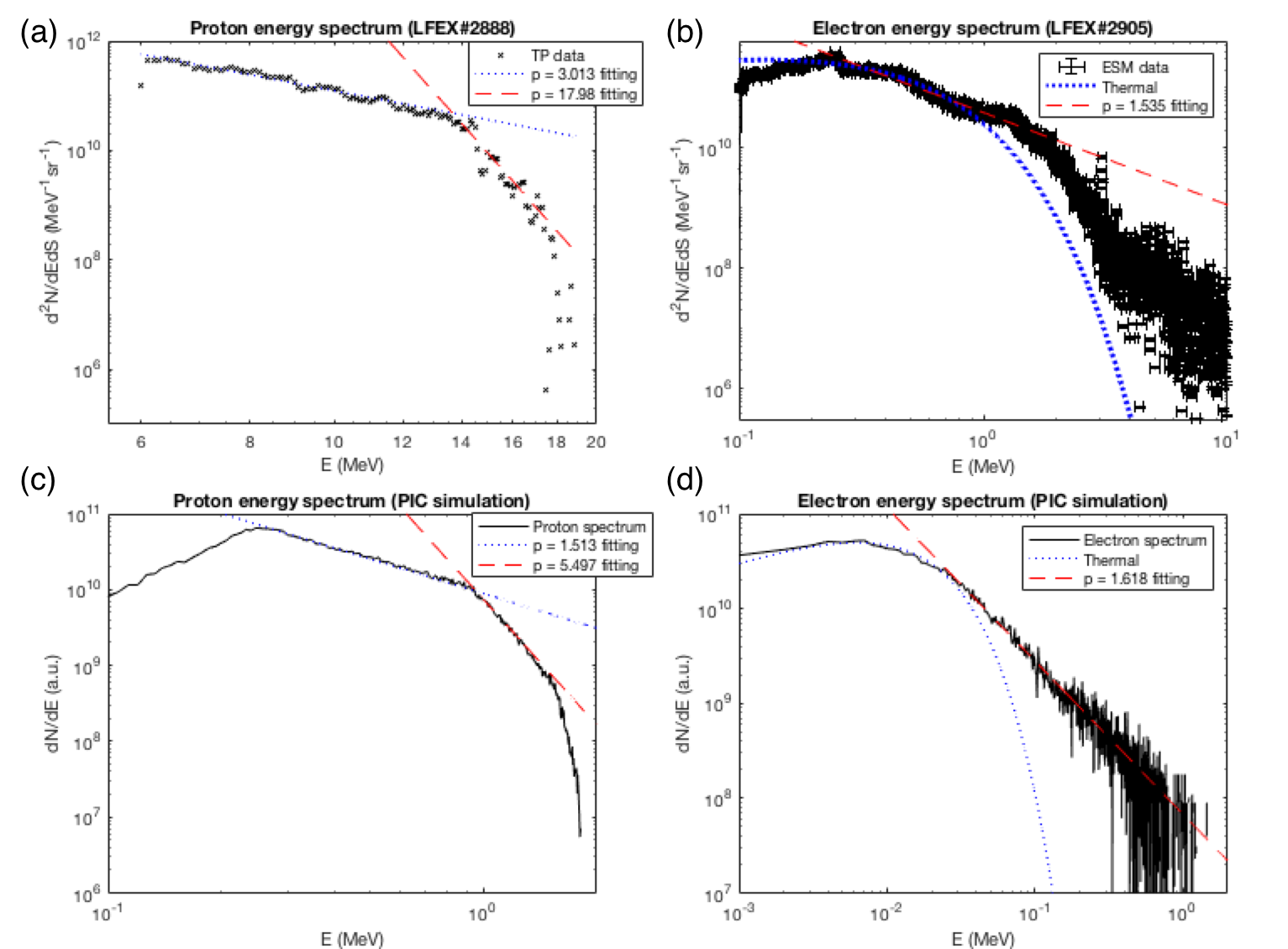}
  \caption{Proton and electron spectra in (a,b) experiment ($l=100$~$\mu$m) and (c,d) PIC simulations ($l=3.3$~$\mu$m). (a) Thomson parabola spectrometer measurement (black crosses) of a proton spectrum with K.E. $>6$~MeV plotted on a logarithmic scale. In the same graph, the best-fitting power-law with slope $p = 3.013$ is plotted (blue dotted line), which has a steep power-law distribution. Near the cutoff energy (about 18 MeV), a steep power-law population with $p=17.98$ (red dashed line) could be fitted. (b) Electron spectrometer measurement (black dots) for electron spectrum with K.E. $>100$~keV on a logarithmic scale. In this spectrum, both thermal and non-thermal components are observed. Maxwell distribution fitting (blue dotted line) and power-law fitting (red dashed line) with slope $p$ = 1.535 are shown in the same graph. From PIC simulation, energy spectra of (c) protons and (d) electrons escaped from the simulation box are plotted. The harder component of proton spectrum (blue dotted line in (c)) is still steeper than the power-law slope of electron (red dashed line in (d)) and both thermal and non-thermal components were observed in the electron spectrum, reproducing the features observed in the experiment.}
  \label{fig:fig3}
\end{figure}

Magnetic reconnection occurs widely in universe\cite{Yamada2010}, making it very natural to perform studies based on telescopic observations of accessible astronomical plasmas. However, the number of cases that can be directly studied based on such plasmas is limited, because most of the consequences of magnetic reconnection occur at locations that are too distant to perform direct measurement. With the need for further study on the fundamental physical processes involved in magnetic reconnection, attempts to observe it in laboratory plasma have been made using different approaches, mainly categorized into magnetic confined plasma\cite{Ono1996} or laser-produced plasma\cite{Zhong2010}. With novel experimental techniques for generating plasma in the laboratory, astrophysical phenomena became attainable in the laboratory environment for direct observation, as long as an appropriate scaling law is applied\cite{Zhong2010}.
\\
\\
Several schemes to observe magnetic reconnection using an intense laser have been demonstrated in previous studies. Zhong \emph{et~al.}\ simulated a solar flare X-ray loop-top source by producing a pair of expanding plasma bubbles around the focus spots of an intense laser, with a frozen-in magnetic field generated by the Biermann battery effect\cite{Zhong2010}. Pei \emph{et~al.}\ observed low-$\beta$ ($\beta = 2\mu_0 nk_B T/B^2$) magnetic reconnection using a magnetic field generated by laser-driven intense current in a double coil\cite{Pei2016}. These experiments are performed with ''long-pulse'' (pulse duration on the order of nanoseconds) lasers having intensities of $10^{14}-10^{16}$~W/cm$^2$. Raymond \emph{et~al.}\ demonstrated magnetic reconnection through a pair of plasma bubbles using a similar method but a relativistic intensity, shorter-pulse laser ($I > 10^{18}$~W/cm$^2$) to achieve a lower $\beta$~value\cite{Raymond2016}.
\\
\\
In the present study, we describe the novel magnetic reconnection scheme, as shown in Fig. \ref{fig:fig1}. The magnetic reconnection experiment is performed using a ''micro-coil'', which was first proposed by Korneev \emph{et~al.}\ as the generation scheme of a sub-gigagauss magnetic field by laser-induced electron acceleration\cite{Korneev2015}. As the laser enters the micro-coil and propagates along the curved surface by multiple reflections at shear incidence\cite{Abe2018},
a current of accelerated electrons is established along the surface\cite{Nakamura2004} ($j_s$). Quasi-instantaneously a return current of thermal electrons $j_r$ sets up, neutralizing the current of super-thermal electrons\cite{Korneev2015}.
With the tendency of charge neutrality, the net current $j_n = j_r+j_s$ always points away from the laser-plasma interaction site, where an electron vacancy is generated due to the electron ejection in the high-energy tail of the electron spectrum. Instead of one-directional $j_n$ [\onlinecite{Korneev2015}], we generated bi-directional $j_n$ in a single micro-coil by simultaneous laser irradiation on two separated sites. This is achieved using a laser beam with a sufficiently large focus spot size, comparable to the opening of the micro-coil. A portion of a laser beam enters from the opening of the micro-coil and multiple reflections are performed, while the remaining part of the laser beam is irradiated on the opposite edge, resulting in an additional $j_n$ pointing in the opposite direction. Based on these processes, a pair of kilo-tesla, anti-parallel magnetic fields, $B_0$, was generated around these bi-directional currents, while expanding plasma was produced along the entire inner surface of the micro-coil by the irradiation of multiple reflected laser beams that entered the micro-coil. As a typical consequence of laser-plasma interaction, the plasma expanded radially inward with the frozen-in anti-parallel magnetic field and converged around center of the micro-coil. This geometrical configuration is prone to make magnetic reconnection occur, as demonstrated by our measurements.
\\
\\
\\
\\
The experiment was performed at the petawatt laser facility LFEX, with maximum 4-beam capability of maximum energy of 500~J per beam at a wavelength of 1,053~nm\cite{Shiraga2011}. In this experiment, two beams of LFEX were operated. One beam of LFEX (pulse duration: 1.2~ps, average energy: 330~J) was focused on the micro-coil with a focus spot having a full width at half maximum (FWHM) of 40 $\mu$m, corresponding to a peak laser intensity of $1.4\times10^{19}$~W/cm$^2$. The micro-coil was fabricated from 10-$\mu$m-thick Cu foil, with a radius of 150 $\mu$m and a length of $l=500$~$\mu$m along its rotational axis. The reconnection magnetic field configuration inside the micro-coil was characterized experimentally by proton deflectometry, a magnetic field diagnosis method by observing and analyzing the deflection of a probing proton beam\cite{Li2006}, which is applicable on kilotesla magnetic field in sub-mm scale\cite{Santos2015, Law2016}. In order to generate the proton beam, another beam of LFEX was focused on an Al foil to perform proton acceleration by the target normal sheath acceleration mechanism\cite{Wilks2001}, with adjustable time difference $\Delta t$ from the micro-coil irradiation, which is provided by the capability of LFEX inter-beam synchronization. Stacks of radiochromic film (RCF) were placed in both the radial and axial directions in order to record the spatial distributions of protons in accelerated reconnection outflow jets. Energy spectra of protons and electrons in the jets were recorded by a Thomson parabola spectrometer (TPS)\cite{Mori2006} placed at one side of the axial direction of the micro-coil, and an electron spectrometer (ESM) was placed at the opposite side.
\\
\\
Before observation of the magnetic reconnection outflow, the magnetic field geometry and amplitude were characterized experimentally. The probing proton beam was deflected under a magnetic field, the deflected pattern was then recorded by a RCF stack placed in the radial direction in respect to the micro-coil axis (see Fig. \ref{fig:fig2}). Monte-Carlo particle tracing simulations, combined with our magnetic field model, showed a correlation between the size of the deflected pattern and the current amplitude, determined the magnetic field inside the micro-coil as $B_0 = 3.0\pm0.3$~kT. The analysis procedure is briefly described as follows. Based on a study by Korneev \emph{et~al.}\cite{Korneev2015}, and further confirmed by PIC simulation, the magnetic field configuration during interaction is correlated with the direction of current flow between the two magnetized regions that carrying the reconnection magnetic field, $B_0$, in opposite directions (indicated as ''Current sheet'' in Fig. \ref{fig:fig1}). From the RCF stack placed in the radial direction, we observed a proton beam accelerated along this current sheet, indicated as "Proton along field boundary" in Fig. \ref{fig:fig2}, and therefore determined the magnetic field geometry. Based on this result, we have sufficient constraints for modeling a magnetic field profile, including both $j_n$ and the current sheet component. Details of this magnetic field analysis are given in the Methods Section.
\\
\\
From two ends of the micro-coil, RCF stacks detected a pair of symmetric proton jets with a maximum K.E. of 6.7~MeV, as shown in Fig. \ref{fig:fig2}, which was interpreted as a typical magnetic reconnection outflow. These proton jets were verified to be a product of magnetic reconnection through PIC simulation, which only appears during simulation conditions in which a bi-directional magnetic field is initially generated and magnetic reconnection occurs. 
In these simulations, the existence of the magnetic reconnection was traced using dissipation measure $\bm{D}_e=\gamma_e[\bm{j} \cdot (\bm{E}+\bm{v}_e\times \bm{B})-\rho_c(\bm{v}_e\cdot \bm{E}))]$, where $\gamma_e$ is the Lorentz factor of bulk electrons, $\bm{j}$ the current density, $\bm{E}$ and $\bm{B}$ the local electric and magnetic field, $\bm{v}_e$ the average velocity of electrons and $\rho_c$ the charge density. This dissipation measure is a Lorentz-invariant scalar quantity of the energy transfer from the electromagnetic field to the plasma in the rest frame of the electron\cite{Zenitani2011}, which is validated in laser-produced plasma\cite{Xu2016}. Magnetic reconnection sites are correlated to the observation of proton jets and bi-directional currents, but not in cases in which neither proton jets nor bi-directional currents are  generated. For example, the contour line of $D_e=2\times10^{26}$ is indicated in Fig. \ref{fig:fig1}(b), which, in our setup configuration, consistent with the expected site for the field reconnection. In this experiment, the source of hydrogen atoms for such a proton jet is a contamination layer with a thickness of nanometers that formed on the surface of the micro-coil, which is a common feature in laser experiments\cite{Allen2004}.
\\
\\
Similar experimental shots were performed, but with shorter micro-coils ($l=100$~$\mu$m). Proton jet was experimentally observed in the axial direction by RCF, with a significantly larger value of the estimated maximum K.E. of 19.6~MeV. In such shorter micro-coils, our 3D-PIC simulations showed that the global magnetic field amplitude did not vary significantly. However, this gives a narrower current $j_n$ through the micro-coil, which is the dominating constraint of the magnetic field geometry. This suppressed the intense current growth inside the electron diffusion region (recognized as region with significant electron dissipation in simulation), which lead to enhancement of the magnetic reconnection rate, the reconnection electric field, and thus an increase in jet particle energy.
\\
\\
The most important result is that power-law distributions with different spectrum steepness were observed in proton and electron energy spectra. As a benefit of this higher maximum proton K.E. in the outflow proton jet, its energy spectrum was clearly measured by TPS (with a K.E. lower limit of 6~MeV for protons), as shown in Fig. \ref{fig:fig3}(a). The maximum K.E. obtained by TPS measurement was 18.8~MeV, which is consistent with RCF stack measurement. The proton spectrum could be well fitted by a power law, $N(E)=N_0 E^{-p}$, with a slope of $p = 3.013$. Similarly, the outflow electron K.E. spectrum was measured by ESM, as shown in Fig. \ref{fig:fig3}(b). Both thermal and non-thermal components could be observed from the electron spectrum, as a benefit of ESM's wider detection range from 100~keV to 100~MeV. Uncertainty of ESM was included in the plot, $\pm$4.5$\%$ on electron energy due to calibration uncertainty and $\pm$5$\%$ on electron number due to detector response uncertainty for electrons. With these uncertainties accounted, power-law fitting on a non-thermal component yielded $p_e = 1.535\pm0.015$, which is a much harder spectrum than that of the protons. Synthetic spectra of particles escaped from PIC simulation box along the jet direction are plotted in Fig. \ref{fig:fig3}(c,d), reproducing the experimental observed features. The cut-off particle energy in simulation is one order lower than experimentally measured spectra, as the consequence of smaller system spatial scale (30 times smaller than real experiment). Similar dependency was confirmed in previous study by 2-D PIC simulation on pair plasmas\cite{Werner2015}.
\\
\\
This difference between the observed hard electron spectrum and the steeper proton power-law spectrum could be explained by the reconnection field direct acceleration process proposed by Melzani \emph{et~al.}\cite{Melzani2014}] where repartition between electron and ions resulted in different orders of magnetization for each species. Here, we estimated the inflow proton and electron magnetization, which is defined as the ratio of energy flux in the reconnecting magnetic field to the energy flux of particles, $\sigma^{hot}_s(B)=B^2/(\mu_0 n_s m_s c^2 \Gamma_s h_{0,s})$, for particle species s, where $n_s$ is the number density of species, $m_s$ the particle mass, $\Gamma_s=(1-v^2_s/c^2)^{-1/2}$ the Lorentz factor, and $h_{0,s}\simeq1+(5/2)(T/mc^2)$ the enthalpy\cite{Melzani2013}.
A two-dimensional, real-scale PIC simulation was performed to estimate the physical quantities $n_s$ and $\Gamma_s h_{0,s}$, where the density profile of the proton plasma layer is set to match the nanometer-scale solid density layer identified in a previous study\cite{Allen2004} in terms of areal density. After accounting for the effect of reduction of dimensionality, parameters ranged between $n_p = n_e = 10^{23}-10^{24}$~m$^{-3}$, $\Gamma_e h_{0,e}=1.5-10$, and $\Gamma_p h_{0,p}\simeq1$. Based on the experimental result of $B_{max}=3.0$~kT, we obtained $\sigma^{hot}_e=40-200$ and $\sigma^{hot}_p=0.05-0.5$. In this range of electron magnetization, $\sigma^{hot}_e$, a hard electron spectrum is expected to be generated through direct acceleration by reconnection electric field, while the proton magnetization is $\sigma^{hot}_p<1$, consistent with the steeper power-law spectrum\cite{Melzani2014}.
\\
\\
\begin{table}[t]
  \caption{Physical parameters in various environments, including the laser experiment of the present study and potential candidates for astronomical plasma with the magnetic reconnection process. The relativistic Alfv\'en velocity $v_A/c=[\sigma_{i+e}/(1+\sigma_{i+e})]^{-1/2}$ is also shown.}
  \begin{tabular}{l|c|c|c|c|c} \hline
    Magnetic reconnection plasma & $B_0$~(T) & $n_e$(m$^{-3}$) & $T_e$(K) & $\sigma^{hot}_e$ & $v_A/c$ \\ \hline
    Laser-driven micro-coil & $3.0\times10^3$ & $10^{23}-10^{24}$ & $10^9-10^{10}$ & $40-200$ & $0.22-0.58$ \\
    Cygnus X-1\cite{DelSanto2013} & $10^3$ & $5\times10^{24}$ & $10^9$ & $130$ & $0.3$ \\
    Microquasar coronae\cite{Reis2013,Melzani2014} & $10^1-10^3$ & $10^{19}-10^{22}$ & $10^9$ & $10^{-1}-10^5$ & $0.003-1$ \\
    Fast reconnection region of GRB jet\cite{Melzani2014,Mckinney2011} & $7\times10^4$ & $10^{16}$ & $10^8$ & $5\times10^{12}$ & $0.9$
  \end{tabular}
\end{table}
In Table 1, we list the results for our laboratory study and some typical astronomical objects, including Cygnus X-1, a well-known example of a galactic X-ray source. For Cygnus X-1, the values of the analytical model \cite{DelSanto2013} are set to $B = 10^3$~T, $n_e = n_p = 5\times10^{22}$~m$^{-3}$, $\Gamma_e h_{0,e}=1.5$, and an electron magnetization of $\sigma^{hot}_e = 130$. Recently, a jet model is proposed as candidate of X-ray emission mechanism, that can account for constraints according to recent detection on Cygnus X-1, as long as hard electron spectra $p=1.4 - 1.5$ are formed in the jets by efficient particle acceleration\cite{Zdziarski2014}. We experimentally showed such a hard electron population could be built in the magnetic reconnection outflow jet, under inflow electron magnetization $\sigma^{cold}_e = 20-100$, which is similar to the electron magnetization of galactic X-ray source Cygnus X-1. According to recent simulation results\cite{Melzani2014}, the power-law slope of accelerated electron depend mainly on $\sigma_e$ and $v_A$, independent of other parameters including simulation spatial scale or time duration. Based on this relation, we would like to emphasize that our result experimentally showed the possibility of hard electron spectrum formation in Cygnus X-1 through magnetic reconnection.
\\
\\
In conclusion, in the present study, the magnetization condition for a typical galactic X-ray source is reproduced in a laboratory magnetic reconnection experiment through generation of a kilotesla-order magnetized relativistic plasma using the micro-coil scheme. In contrast to previous attempts to study astrophysical phenomena by applying scaling law on laboratory plasma, most of the plasma parameters in our experiment are directly comparable to that of Cygnus X-1. Main differences are the spatial and temporal scales, which are reflected in the difference of cut-off electron energy $\gamma_{max} mc^2$ between our experiment and Cygnus X-1 jet model\cite{Zdziarski2014}: $\gamma_{max}\sim20$ was detected in our experiment in a system of 100~$\mu$m ($10^{-4}$~m), while $\gamma_{max}\sim10^6$ was modeled in Cygnus X-1 jets at height $h=7\times10^9$~m from plane of binary plane. With the fact that the unknown spatial scale of magnetic reconnection region contributed to the outflowing particle energy must be shorter than this height, the $10^{14}$ difference in system scale do not contradict to the $10^5$ difference in cut-off energy. Spatial scale of magnetic reconnection region in Cygnus X-1 could be estimated, if further studies are performed on this scaling in electron-ion plasma in future. Nevertheless, the electron magnetization and the associated reconnection dynamics between our experiment and Cygnus X-1 are directly comparable. Therefore our results experimentally showed the possibility of magnetic reconnection contribution to X-ray emission from galactic X-ray source, such as Cygnus X-1, through Compton upscattering of stellar blackbody photons as Zdziarski \emph{et~al.}\cite{Zdziarski2014} suggested.

\section{Method}
\subsection{Three-dimensional particle-in-cell simulation}
\begin{figure}[htbp]
\centering\includegraphics[width=\linewidth]{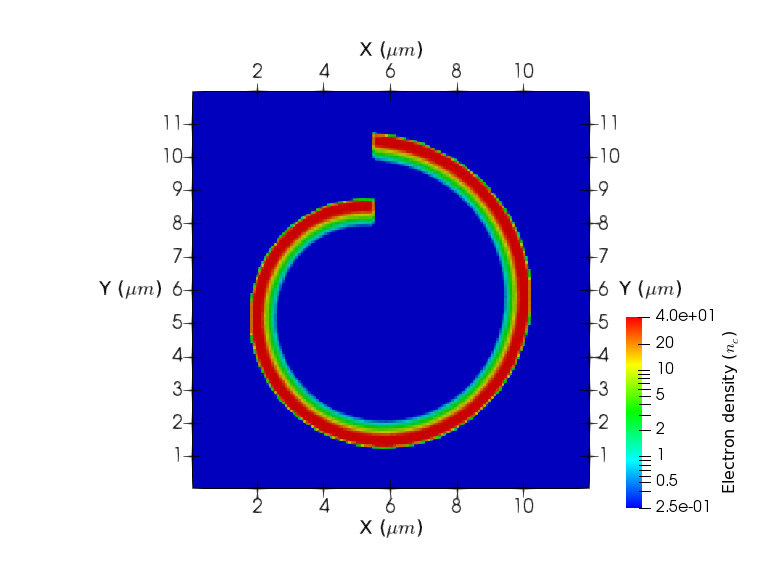}
    \caption{(Extended) Initial electron density profile of PIC simulation plotted on the x-y plane.}
    \label{fig:exfig1}
\end{figure}

Numerical simulations of the laser-plasma interaction and its resulting electromagnetic field and particle kinetics were performed by a three-dimensional particle-in-cell (PIC) simulation code EPOCH\cite{Arber2015}. In these simulations, plasma representing the micro-coil is initially configured by the target shape analytically defined\cite{Korneev2015}: $r(\theta)=r_0[1+(\delta r/r_0)(\theta/2\pi)]$.
\\
\\
Limited by our available computational resources, the target size in our simulation is 1/30 of the real target in all dimensions, $r_0 = 3$~$\mu$m and $\delta r=2$~$\mu$m. Moreover, collision effects are not included in this simulation. The simulation box is $300\times300\times1,250$~cells in Cartesian coordinates, corresponding to $12\times12\times50$~$\mu$m, and a single time step is 0.059~fs. The coordinate system is defined as follows. The cross section of the micro-coil is on the $x$-$y$ plane, with the x-axis parallel to the incident laser propagation direction, and the y-axis parallel to its polarization. The $z$-axis is the axial direction of the micro-coil, as well as the main direction of the reconnection magnetic field. The micro-coil cross-section density profile is shown in Fig. \ref{fig:exfig1} (Extended Data). In addition, $r(\theta)$ is defined in polar coordinates with the origin set at $x=y=5.5$~$\mu$m, where $\theta = 0$ is in the $+y$ direction, and $\theta$ increases in the anti-clockwise direction.
\\
\\
The target consists of two layers. An outer layer representing the bulk of the micro-coil, and an inner layer representing the pre-plasma generated by the nanosecond-order pre-pulse before the arrival of the main pulse. The outer surface consists of fully ionized copper plasma with an ion density of $n_i=1.5\times10^{21}$~cm$^{-3}$, one ion, and 29 electrons per cell. The inner layer consists of proton plasma with an exponential density profile of $n_p=n_e=40n_c\cdot e^{-d/\tau}$, where $n_c=1.01\times10^{21}$~cm$^{-3}$ is the electron critical density, d represents the distance from the outer layer, and $\tau=0.1$~$\mu$m. In the proton plasma, five protons and five electrons are placed per cell. The real ion-electron mass ratio is used in this simulation. This density profile is uniform along the z-direction, with a finite length of from 3.3 to 16.7 $\mu$m, representing different types of micro-coils in the experiment.
\\
\\
In the simulation, the incident laser entered the simulation box from the $-x$ side boundary. For the maximum laser intensity, $I_0=1.0\times10^{19}$~W/cm$^2$ at a wavelength of $\lambda_0=1.05$~$\mu$m, a FWHM of the Gaussian intensity distribution of 1.33~$\mu$m, a temporal Gaussian pulse at FWHM of 1.2~ps, and a peak intensity at $t=0.75$~ps. Polarization of the incident laser is p-polarization, which is the actual experimental condition. In order to separate out the incidence laser field, electromagnetic field snapshots are averaged for a single laser period.
\\
\\
\subsection{Magnetic field analysis}
\begin{figure}[htbp]
  \centering\includegraphics[width=\linewidth]{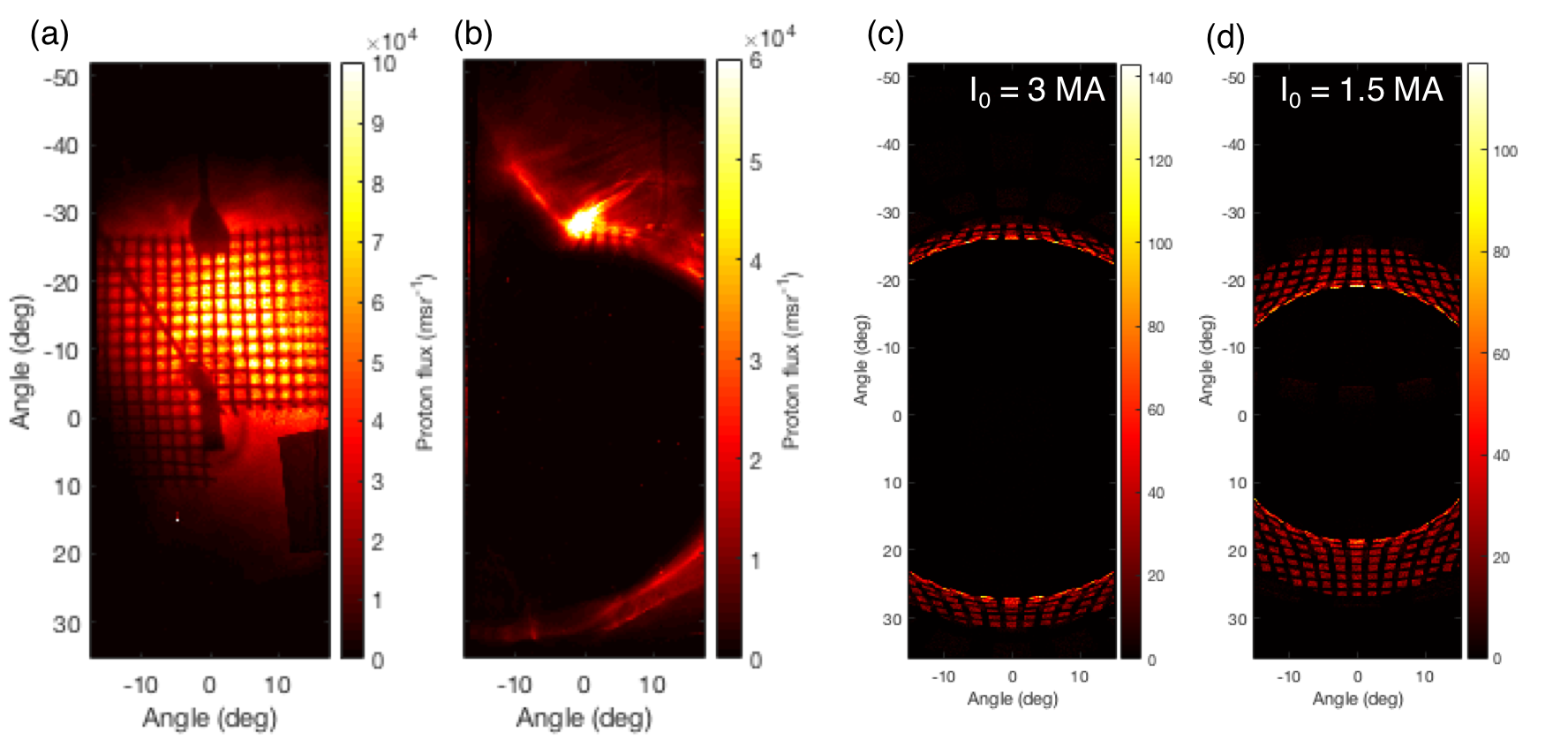}
  \caption{(Extended) (a,b) Proton proton beam spatial pattern obtained in experiment and (c,d) Simulated proton beam pattern reproduced by Monte-Carlo simulation with magnetic field modeled by different $I_0$. Proton pattern (a) without magnetic field deflection and (b) under magnetic field deflection are shown, which probed the magnetic field generated by the micro-coil. In both cases, a grid is placed between the proton source and micro-coil. Under same current model, different value of $I_0$ produces different magnetic field amplitude. In (c) we showed simulated proton pattern of the best-fit pattern, when $I_0=3$~MA, and another simulated proton pattern when $I_0=1.5$~MA, a less-intense magnetic field configuration assumed, is shown in (d). Size of the void pattern increases with the value $I_0$, which is directly related to the magnetic field strength $B_0$.}
  \label{fig:exfig2}
\end{figure}

\begin{figure}[htbp]
\includegraphics[width=\linewidth]{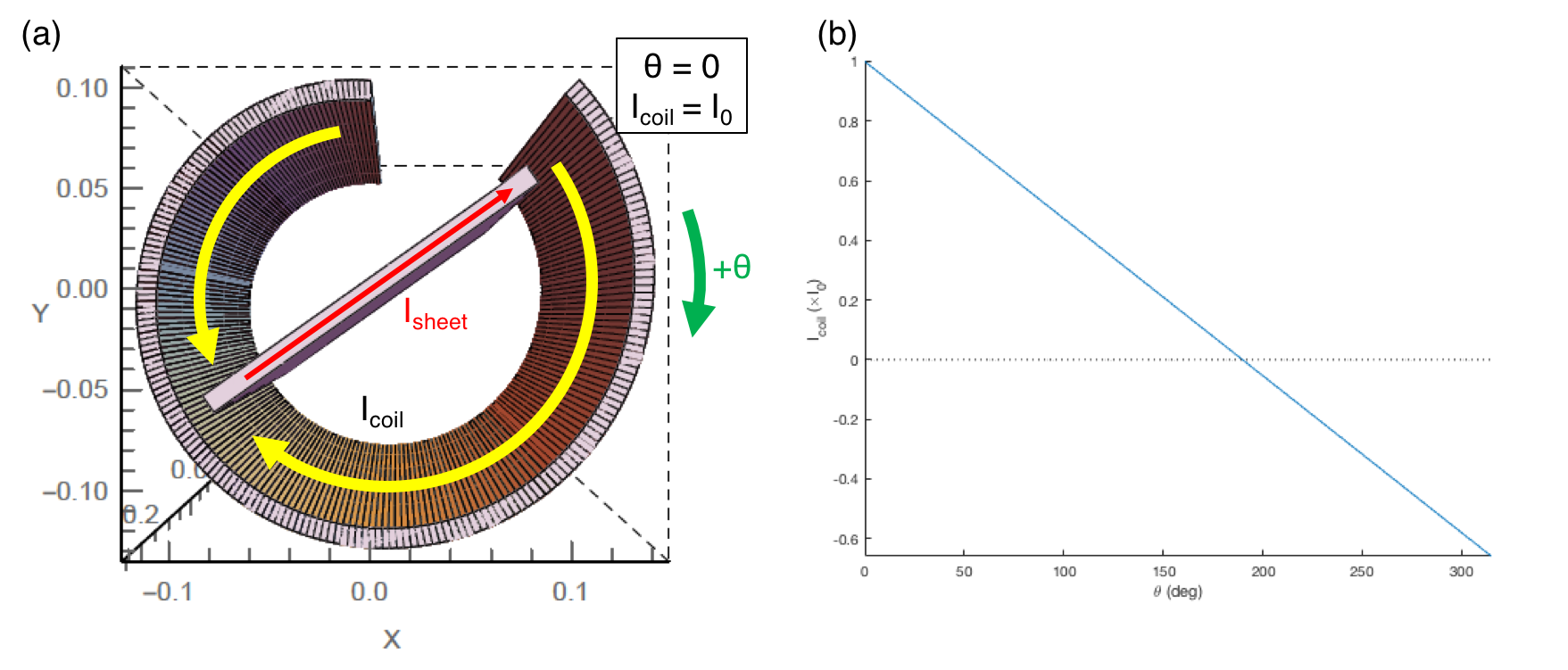}
    \caption{(Extended) Current model constructed based on experimental result. (a) Schematic of current model, consisted of current flow in $\theta$ direction along the micro-coil $I_{coil}$ and current sheet $I_{sheet}$ across the micro-coil. Positive value of $I_{coil}$ corresponds to clockwise current flow in diagram. (b) Distribution of $I_{coil}$ along $\theta$ direction. $\theta=0$ correspond to one end of $I_{coil}$, which is the incident laser focus spot in experiment.}
    \label{fig:exfig3}
\end{figure}

\begin{figure}[htbp]
\includegraphics[width=\linewidth]{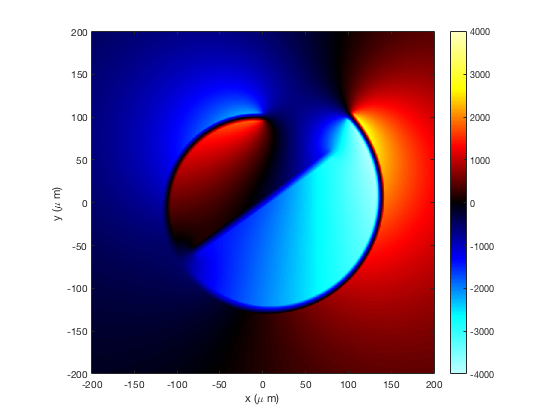}
    \caption{(Extended) Cross section of B-field map determined by proton probe beam deflection followed by Monte-Carlo simulation analysis.}
    \label{fig:exfig4}
\end{figure}

The deflected protons were recorded by an RCF stack consisting of a number of RCFs and aluminum filters positioned 3~cm from the micro-coil. The signal detected by the RCFs depends solely on the particle energy deposition within the thin ($7-15$~$\mu$m) active layer Therefore, the spatial pattern of the signal on a single RCF can be reasonably assumed to be dominated by protons carrying K.E. that correspond to the Bragg peak on the active layer. The value of K.E. corresponding to each RCF layer is calculated by the PHITS Monte Carlo particle transport simulation code\cite{Sato2018}.
\\
\\
Here, we refer to a single RCF layer corresponding to a proton K.E. of 8.5~MeV, which was chosen such that the deflected pattern was expected to be most clearly observed. The corresponding proton velocity, $v_p=4.0\times10^7$~ms$^{-1}$, has a time of flight of $t=l/v_p=75$~ps for $l=3$~mm, which is the distance between the proton generation site and the micro-coil. We then set the LFEX inter-beam time difference such that time difference between laser irradiation of two targets is $\delta t = 69$~ps in order to capture the magnetic field generated during and after the laser irradiation. The unaffected proton beam pattern and deflected proton beam pattern are shown in Fig. \ref{fig:exfig2} (Extended Data).
\\
\\
In order to determine the magnetic field from a single deflected proton pattern, the magnetic field geometry must be assumed. Here, we constructed a current model assuming that the magnetic field is generated by current flow along the micro-coil and the current sheet (indicated by green and purple arrows, respectively, in Fig. \ref{fig:fig1}). In this model, the micro-coil is modeled as the analytical defined shape $r(\theta)=r_0[1+(\delta r/r_0)(\theta/2\pi)]$, with $r_0=100$~$\mu$m, $\delta r = 50$~$\mu$m. Schematic diagram of the current model is shown in Fig. \ref{fig:exfig3}(a). The current density is assumed to be uniform along the z-direction, and the extension length is assumed to be the same as the length of the micro-coil (500~$\mu$m). Based on our PIC simulation results, the current flow along the micro-coil could be properly modeled as varying linearly along the $\theta$ direction with two different current density values at the ends of the micro-coil. The modeled coil current distribution is plotted in Fig. \ref{fig:exfig3}(b). The current sheet is modeled at the center of the micro-coil.
\\
\\
In this current model, the orientation of the current sheet and the current density value at micro-coil ends must be assumed. As we observed from PIC simulation results, the tangential current density at the intersection between the current sheet and the micro-coil is close to zero. This is used as an additional constraint, and our current model is well defined as long as the current sheet direction is known. As mentioned in the Main section, this direction was experimentally determined by measuring the proton beam direction, which is 145$^\circ$ from the incidence laser direction. Based on this constraint, the current sheet is in parallel to this direction, and the current amplitudes at the ends are $I_0$ and $-0.66I_0$. In the simulation, the amplitude of the current sheet is assumed constant across the micro-coil.
\\
\\
With this magnetic field model depending only on the value of $I_0$ and current sheet, classical Runge-Kutta scheme Monte-Carlo simulations are performed on particle tracing in the proton probing beam under the Lorentz force by the magnetic field. Magnetic field profile is calculated from the above current model by 3-D magnetostatic code RADIA\cite{Chubar1998}. Divergence angle of source protons are obtained from experimental measurement of the undeflected proton beam. In simulation, source protons are assumed to be monoenergetic because of the Bragg peak property of proton energy deposition on RCF active layer, as discussed above. We performed a parameter scan on $I_0$ and current sheet $I_{sheet} = cI_0$ and found that proton pattern can only be reproduced when $0<c<0.3$. Within this range of c, $I_0=2.49 - 2.86$~MA best fitted the experimental result, and $B_{max}=3.0\pm0.3$~kT is determined with $10\%$ uncertainty due to the unknown $I_{sheet}$. A magnetic field profile on the x-y plane at $z=0$ with $c=0.2$ is shown in Fig. \ref{fig:exfig4} (Extended Data).

\section{Supplementary Information}
\subsection{Bi-directional current generation in the micro-coil}
\begin{figure}[htbp]
\centering\includegraphics[width=0.45\linewidth]{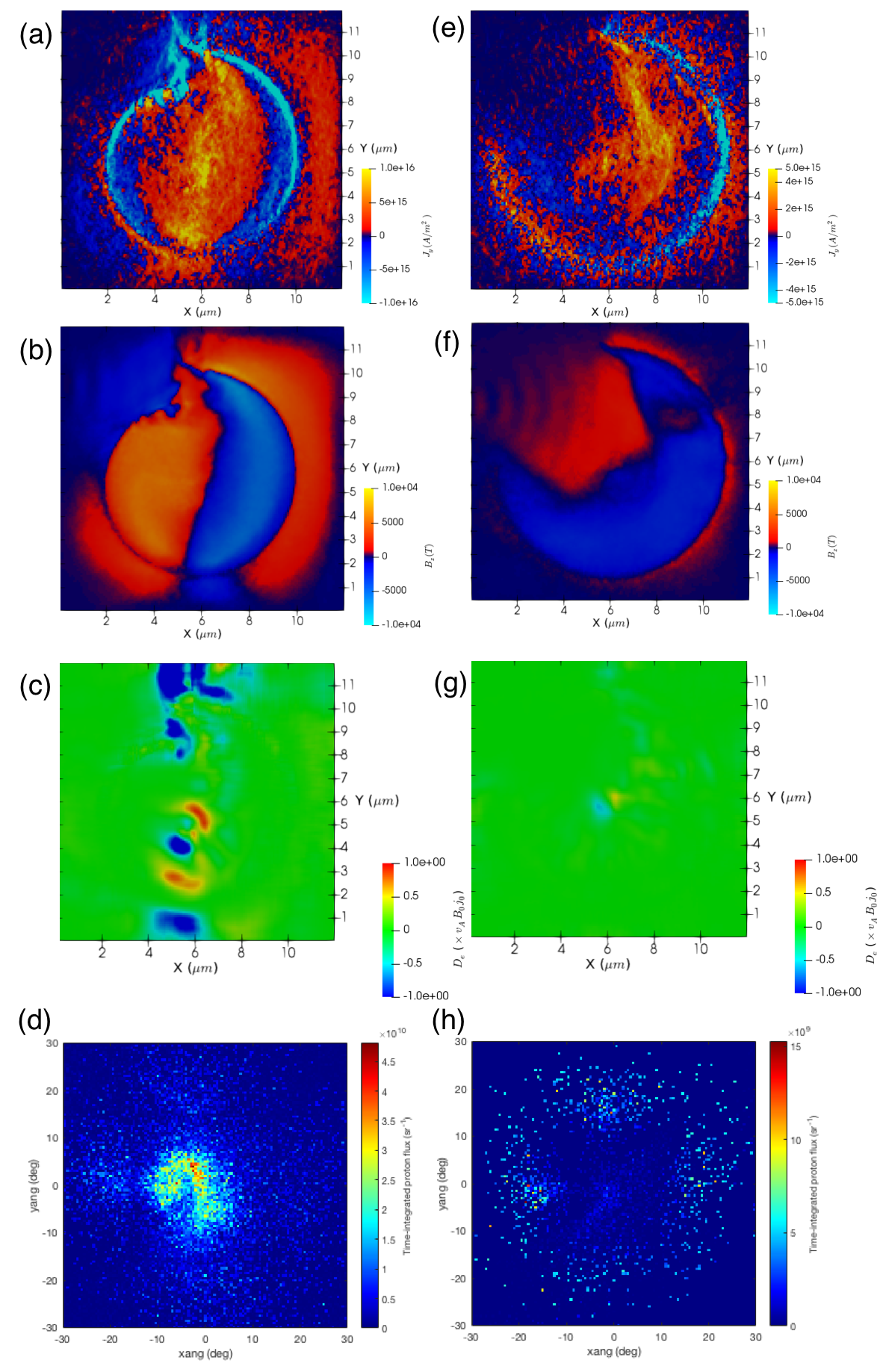}
  \caption{(Extended) Results for the (a)-(d) micro-coil and (e)-(h) open-cylinder cases. (a) and (e) Current density $J_y$ at $z=0$. An intense $-y$ component on two sides was observed for the micro-coil case, indicating bi-directional current generation. For the open-cylinder case, reversal of the sign of the y-component was observed at the lower end, and bi-directional current was not generated. (b) and (f) Magnetic field $B_z$ at $z=0$. Here, $B_z$ in opposite directions is observed in both cases, despite the difference in current generation. (c) and (g) Dissipation measure $D_e$ at $z=0$, normalized by $v_{A0} B_0 j_0$, with $v_{A0} = 0.023 c$, $B_0 = 10^4$~T and $j_0=10^{16}$~A/m$^2$ the typical values of Alfv\'en velocity, magnetic field and current density respectively. Efficient energy transfer from the electromagnetic field to plasmas was observed only in the micro-coil case, where bi-directional $B_z$ was generated by the initial bi-directional current and then reconnected. (d) and (h) Reconnection outflow proton angular distribution. The momentum vector direction distribution of the proton that escaped from the simulation box across the $-z$ boundary is plotted, while the $+z$ side showed similar results. All protons with K.E. $>400$~keV are integrated, and the origin of the plot indicates the direction parallel to the z-axis. The outflow proton beam is only observed in the micro-coil case, as a consequence of magnetic reconnection.}
  \label{fig:exfig5}
\end{figure}

From the three-dimensional PIC simulation described in the Method section, we found that laser irradiation of both ends of the micro-coil is the condition for bi-directional current generation, which initiated magnetic reconnection inside the micro-coil in our experiment. We performed PIC simulations for two different cases. Case 1 (micro-coil case) is identical to the simulation mentioned in the Method section, and Case 2 (open-cylinder case) is a control simulation, where the initial plasma geometry is modified to an open cylinder, which is three-quarters of a cylinder, so that laser irradiation of the lower end is significantly reduced. The simulation results are shown in Fig. \ref{fig:exfig5} (Extended Data). Based on these results, we observed that bi-directional current is generated only in the micro-coil case, where laser irradiated both ends of the cylinder. As a result, dissipation from the electromagnetic field to the plasma is much more efficient in the micro-coil case, even though anti-parallel $B_z$ is observed in both cases. Outflow proton jets, as shown in Figs. \ref{fig:exfig5}(d) and  \ref{fig:exfig5}(h) (Extended Data), is observed only in the micro-coil case, in which magnetic reconnection occurs. This bi-directional current generation was not observed in previous simulation result\cite{Korneev2015}, which may be a consequence of the artificially massive ($m_i=3,940 m_p$) outer sublayer or the different configuration of the incidence laser.

\subsection{Magnetic field modeling based on proton beam self-emission measurement}
\begin{figure}[htbp]
\centering\includegraphics[width=\linewidth]{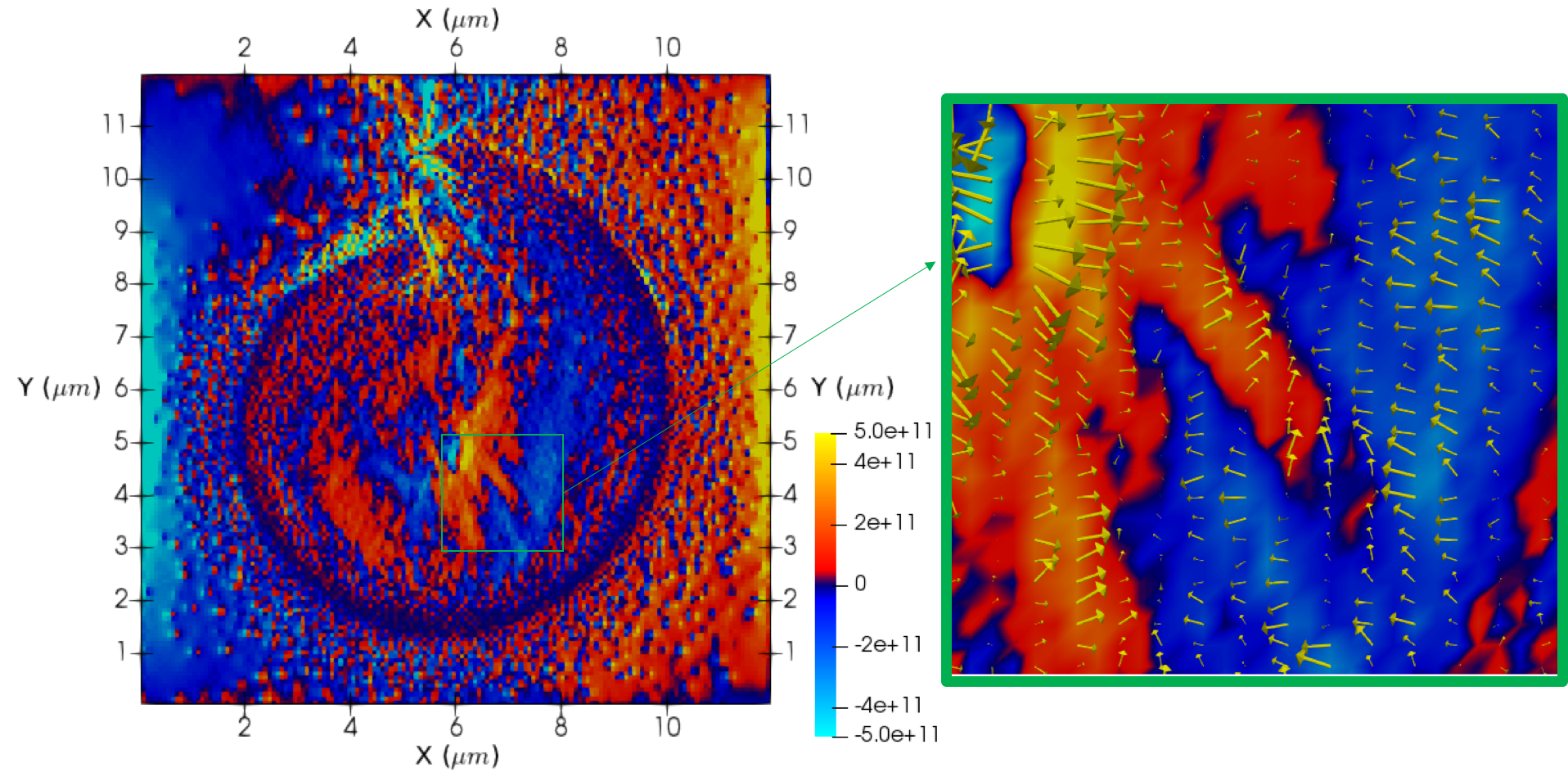}
    \caption{(Extended) Electric field $E_x$ at $z=0$. Vectors of electric field x-y component are plotted in region of green box, showed how the electric field contributed on the proton flow along the reconnection field boundary.}
    \label{fig:exfig6}
\end{figure}

\begin{figure}[htbp]
\centering\includegraphics[width=\linewidth]{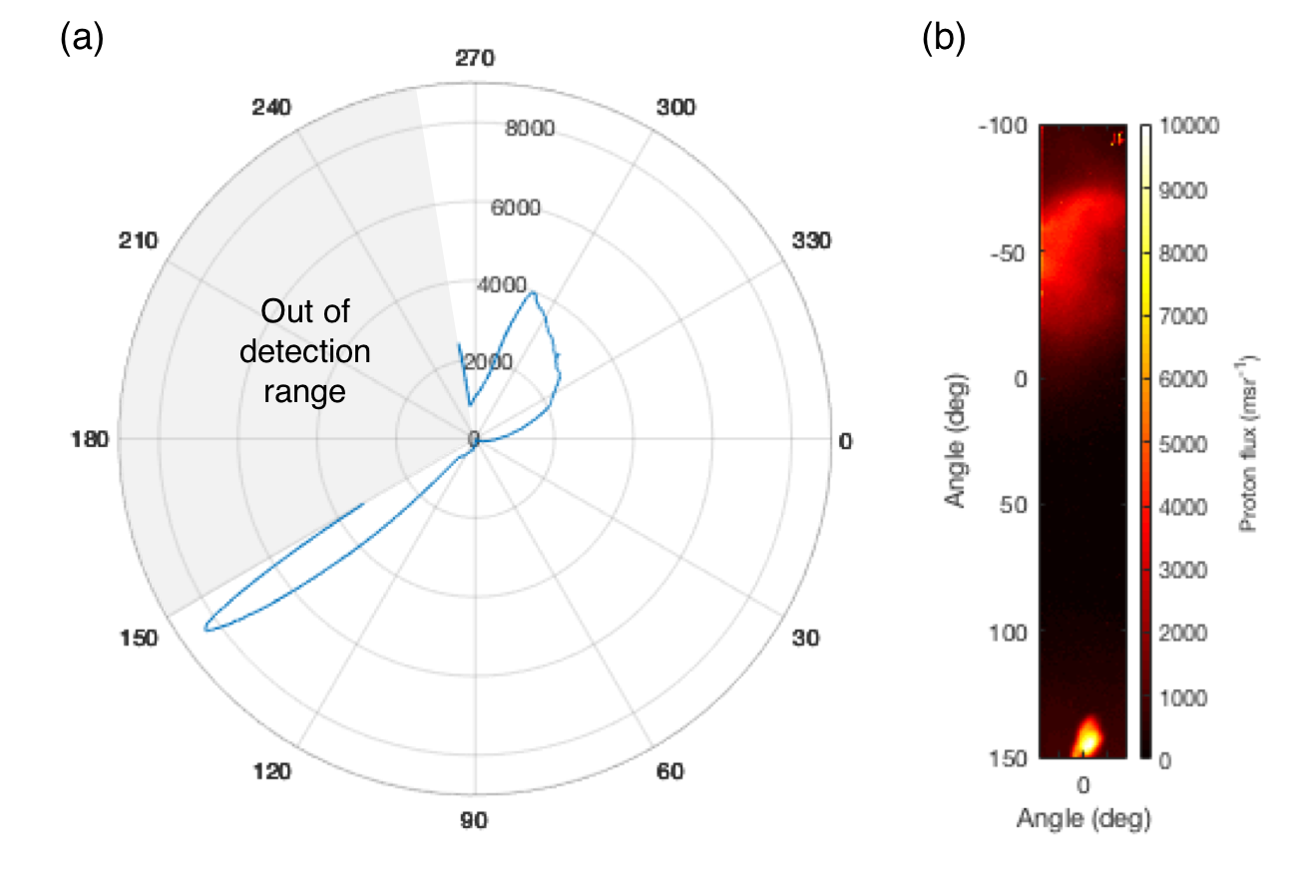}
    \caption{(Extended) Distribution of protons detected in the radial direction of the micro-coil. (a) Polar plot of average RCF signal from -5$^{\circ}$ to 5$^{\circ}$. (b) RCF signal obtained in experiment. The highly collimated signal observed at 145$^\circ$ is a typical signature of proton beam. The direction of this proton beam provided information of the magnetic field geometry inside the micro-coil.}
    \label{fig:exfig7}
\end{figure}

In the previous study\cite{Korneev2015}, the formation of current flow between the anti-parallel magnetic field regions was identified as a consequence of fast electrons confined by the kilotesla-order magnetic field. Based on Fig. \ref{fig:exfig5}(a) (Extended Data), a similar current structure is observed at the center of the micro-coil. At the same time, this confinement decouples electrons from ions, which generates an electric field that points into the current flow. The x-component of the electric field $E_x$ on the x-z plane is shown in Fig. \ref{fig:exfig6} (Extended Data), which clearly shows this feature around the current sheet. As a result of this effect, a fraction of protons (fewer than the number of electrons) is guided along the current flow, so that the net current is dominated by electron flow. Through this mechanism, protons are guided along the magnetic field boundary and escape from the micro-coil as a proton beam along the direction parallel to this boundary.
\\
\\
In the experiment, this proton beam is detected by an RCF stack, and the results are shown in Fig. \ref{fig:exfig7} (Extended Data). The proton flux has a single peak at 145$^\circ$ from the incidence laser direction. In our magnetic field model, we approximated the magnetic field configuration with a boundary aligned along this direction. Moreover, at the intersection between the current sheet and micro-coil, the tangential component of current density changes its sign where its magnitude falls to zero, as shown in Fig. \ref{fig:exfig5} (Extended Data). This feature is included as an assumption in the magnetic field model.

\bibliographystyle{naturemag}
\bibliography{2018MRX}

\begin{thebibliography}{10}
\expandafter\ifx\csname url\endcsname\relax
  \def\url#1{\texttt{#1}}\fi
\expandafter\ifx\csname urlprefix\endcsname\relax\def\urlprefix{URL }\fi
\providecommand{\bibinfo}[2]{#2}
\providecommand{\eprint}[2][]{\url{#2}}

\bibitem{Yamada2010}
\bibinfo{author}{Yamada, M.}, \bibinfo{author}{Kulsrud, R.} \&
  \bibinfo{author}{Ji, H.}
\newblock \bibinfo{title}{{Magnetic reconnection}}.
\newblock \emph{\bibinfo{journal}{Reviews of Modern Physics}}
  \textbf{\bibinfo{volume}{82}}, \bibinfo{pages}{603--664}
  (\bibinfo{year}{2010}).
\newblock \eprint{arXiv:0708.1752}.

\bibitem{Sweet1958}
\bibinfo{author}{Sweet, P.~A.}
\newblock \bibinfo{title}{14. the neutral point theory of solar flares}.
\newblock In \emph{\bibinfo{booktitle}{Symposium-International Astronomical
  Union}}, vol.~\bibinfo{volume}{6}, \bibinfo{pages}{123--134}
  (\bibinfo{organization}{Cambridge University Press}, \bibinfo{year}{1958}).

\bibitem{Hughes1995}
\bibinfo{author}{Hughes, W.~J.}
\newblock \bibinfo{title}{The magnetopause, magnetotail, and magnetic
  reconnection}.
\newblock \emph{\bibinfo{journal}{Introduction to Space Physics}}
  \bibinfo{pages}{227--287} (\bibinfo{year}{1995}).
\newblock \urlprefix\url{https://ci.nii.ac.jp/naid/10003737051/}.

\bibitem{Di1998}
\bibinfo{author}{Di~Matteo, T.}
\newblock \bibinfo{title}{Magnetic reconnection: flares and coronal heating in
  active galactic nuclei}.
\newblock \emph{\bibinfo{journal}{Monthly Notices of the Royal Astronomical
  Society}} \textbf{\bibinfo{volume}{299}}, \bibinfo{pages}{L15--L20}
  (\bibinfo{year}{1998}).

\bibitem{Zdziarski2014}
\bibinfo{author}{Zdziarski, A.~A.}, \bibinfo{author}{Pjanka, P.},
  \bibinfo{author}{Sikora, M.} \& \bibinfo{author}{Stawarz, {\L}.}
\newblock \bibinfo{title}{Jet contributions to the broad-band spectrum of cyg
  x-1 in the hard state}.
\newblock \emph{\bibinfo{journal}{Monthly Notices of the Royal Astronomical
  Society}} \textbf{\bibinfo{volume}{442}}, \bibinfo{pages}{3243--3255}
  (\bibinfo{year}{2014}).

\bibitem{Ono1996}
\bibinfo{author}{Ono, Y.}, \bibinfo{author}{Yamada, M.}, \bibinfo{author}{Akao,
  T.}, \bibinfo{author}{Tajima, T.} \& \bibinfo{author}{Matsumoto, R.}
\newblock \bibinfo{title}{Ion acceleration and direct ion heating in
  three-component magnetic reconnection}.
\newblock \emph{\bibinfo{journal}{Physical review letters}}
  \textbf{\bibinfo{volume}{76}}, \bibinfo{pages}{3328} (\bibinfo{year}{1996}).

\bibitem{Zhong2010}
\bibinfo{author}{Zhong, J.} \emph{et~al.}
\newblock \bibinfo{title}{{Modelling loop-top X-ray source and reconnection
  outflows in solar flares with intense lasers}}.
\newblock \emph{\bibinfo{journal}{Nature Physics}}
  \textbf{\bibinfo{volume}{6}}, \bibinfo{pages}{984--987}
  (\bibinfo{year}{2010}).
\newblock \urlprefix\url{http://dx.doi.org/10.1038/nphys1790}.

\bibitem{Pei2016}
\bibinfo{author}{Pei, X.~X.} \emph{et~al.}
\newblock \bibinfo{title}{{Magnetic reconnection driven by Gekko XII lasers
  with a Helmholtz capacitor-coil target}}.
\newblock \emph{\bibinfo{journal}{Physics of Plasmas}}
  \textbf{\bibinfo{volume}{23}} (\bibinfo{year}{2016}).

\bibitem{Raymond2016}
\bibinfo{author}{Raymond, A.} \emph{et~al.}
\newblock \bibinfo{title}{Relativistic magnetic reconnection in the
  laboratory}.
\newblock \emph{\bibinfo{journal}{arXiv preprint arXiv:1610.06866}}
  (\bibinfo{year}{2016}).

\bibitem{Korneev2015}
\bibinfo{author}{Korneev, P.}, \bibinfo{author}{D'Humi{\`{e}}res, E.} \&
  \bibinfo{author}{Tikhonchuk, V.}
\newblock \bibinfo{title}{{Gigagauss-scale quasistatic magnetic field
  generation in a snail-shaped target}}.
\newblock \emph{\bibinfo{journal}{Physical Review E}}
  \textbf{\bibinfo{volume}{91}}, \bibinfo{pages}{043107}
  (\bibinfo{year}{2015}).
\newblock \urlprefix\url{http://link.aps.org/doi/10.1103/PhysRevE.91.043107}.

\bibitem{Abe2018}
\bibinfo{author}{Abe, Y.} \emph{et~al.}
\newblock \bibinfo{title}{Whispering gallery effect in relativistic optics}.
\newblock \emph{\bibinfo{journal}{JETP Letters}}
  \textbf{\bibinfo{volume}{107}}, \bibinfo{pages}{351--354}
  (\bibinfo{year}{2018}).

\bibitem{Nakamura2004}
\bibinfo{author}{Nakamura, T.}, \bibinfo{author}{Kato, S.},
  \bibinfo{author}{Nagatomo, H.} \& \bibinfo{author}{Mima, K.}
\newblock \bibinfo{title}{Surface-magnetic-field and fast-electron
  current-layer formation by ultraintense laser irradiation}.
\newblock \emph{\bibinfo{journal}{Phys. Rev. Lett.}}
  \textbf{\bibinfo{volume}{93}}, \bibinfo{pages}{265002}
  (\bibinfo{year}{2004}).
\newblock
  \urlprefix\url{https://link.aps.org/doi/10.1103/PhysRevLett.93.265002}.

\bibitem{Zenitani2011}
\bibinfo{author}{Zenitani, S.}, \bibinfo{author}{Hesse, M.},
  \bibinfo{author}{Klimas, A.} \& \bibinfo{author}{Kuznetsova, M.}
\newblock \bibinfo{title}{New measure of the dissipation region in
  collisionless magnetic reconnection}.
\newblock \emph{\bibinfo{journal}{Physical review letters}}
  \textbf{\bibinfo{volume}{106}}, \bibinfo{pages}{195003}
  (\bibinfo{year}{2011}).

\bibitem{Shiraga2011}
\bibinfo{author}{Shiraga, H.} \emph{et~al.}
\newblock \bibinfo{title}{Fast ignition integrated experiments with gekko and
  {LFEX} lasers}.
\newblock \emph{\bibinfo{journal}{Plasma Physics and Controlled Fusion}}
  \textbf{\bibinfo{volume}{53}}, \bibinfo{pages}{124029}
  (\bibinfo{year}{2011}).
\newblock
  \urlprefix\url{https://doi.org/10.1088%2F0741-3335%2F53%2F12%2F124029}.

\bibitem{Li2006}
\bibinfo{author}{Li, C.} \emph{et~al.}
\newblock \bibinfo{title}{Measuring e and b fields in laser-produced plasmas
  with monoenergetic proton radiography}.
\newblock \emph{\bibinfo{journal}{Physical review letters}}
  \textbf{\bibinfo{volume}{97}}, \bibinfo{pages}{135003}
  (\bibinfo{year}{2006}).

\bibitem{Santos2015}
\bibinfo{author}{Santos, J.} \emph{et~al.}
\newblock \bibinfo{title}{Laser-driven platform for generation and
  characterization of strong quasi-static magnetic fields}.
\newblock \emph{\bibinfo{journal}{New Journal of Physics}}
  \textbf{\bibinfo{volume}{17}}, \bibinfo{pages}{083051}
  (\bibinfo{year}{2015}).

\bibitem{Law2016}
\bibinfo{author}{Law, K.} \emph{et~al.}
\newblock \bibinfo{title}{Direct measurement of kilo-tesla level magnetic field
  generated with laser-driven capacitor-coil target by proton deflectometry}.
\newblock \emph{\bibinfo{journal}{Applied Physics Letters}}
  \textbf{\bibinfo{volume}{108}}, \bibinfo{pages}{091104}
  (\bibinfo{year}{2016}).

\bibitem{Wilks2001}
\bibinfo{author}{Wilks, S.~C.} \emph{et~al.}
\newblock \bibinfo{title}{{Energetic proton generation in ultra-intense
  laser--solid interactions}}.
\newblock \emph{\bibinfo{journal}{Physics of Plasmas}}
  \textbf{\bibinfo{volume}{8}}, \bibinfo{pages}{542} (\bibinfo{year}{2001}).
\newblock
  \urlprefix\url{http://scitation.aip.org/content/aip/journal/pop/8/2/10.1063/1.1333697}.

\bibitem{Mori2006}
\bibinfo{author}{Mori, M.} \emph{et~al.}
\newblock \bibinfo{title}{{New Detection Device for Thomson Parabola
  Spectrometer for Diagnosis of the Laser-Plasma Ion Beam}}.
\newblock \emph{\bibinfo{journal}{Plasma and Fusion Research}}
  \textbf{\bibinfo{volume}{1}}, \bibinfo{pages}{042} (\bibinfo{year}{2006}).

\bibitem{Xu2016}
\bibinfo{author}{Xu, Z.} \emph{et~al.}
\newblock \bibinfo{title}{Characterization of magnetic reconnection in the
  high-energy-density regime}.
\newblock \emph{\bibinfo{journal}{Physical Review E}}
  \textbf{\bibinfo{volume}{93}}, \bibinfo{pages}{033206}
  (\bibinfo{year}{2016}).

\bibitem{Allen2004}
\bibinfo{author}{Allen, M.} \emph{et~al.}
\newblock \bibinfo{title}{{Direct Experimental Evidence of Back-Surface Ion
  Acceleration from Laser-Irradiated Gold Foils}}.
\newblock \emph{\bibinfo{journal}{Physical Review Letters}}
  \textbf{\bibinfo{volume}{93}}, \bibinfo{pages}{265004}
  (\bibinfo{year}{2004}).
\newblock
  \urlprefix\url{https://link.aps.org/doi/10.1103/PhysRevLett.93.265004}.

\bibitem{Werner2015}
\bibinfo{author}{Werner, G.}, \bibinfo{author}{Uzdensky, D.},
  \bibinfo{author}{Cerutti, B.}, \bibinfo{author}{Nalewajko, K.} \&
  \bibinfo{author}{Begelman, M.}
\newblock \bibinfo{title}{The extent of power-law energy spectra in
  collisionless relativistic magnetic reconnection in pair plasmas}.
\newblock \emph{\bibinfo{journal}{The Astrophysical Journal Letters}}
  \textbf{\bibinfo{volume}{816}}, \bibinfo{pages}{L8} (\bibinfo{year}{2015}).

\bibitem{Melzani2014}
\bibinfo{author}{Melzani, M.}, \bibinfo{author}{Walder, R.},
  \bibinfo{author}{Folini, D.}, \bibinfo{author}{Winisdoerffer, C.} \&
  \bibinfo{author}{Favre, J.~M.}
\newblock \bibinfo{title}{{The energetics of relativistic magnetic
  reconnection: ion-electron repartition and particle distribution hardness}}.
\newblock \emph{\bibinfo{journal}{Astronomy {\&} Astrophysics}}
  \textbf{\bibinfo{volume}{570}}, \bibinfo{pages}{A112} (\bibinfo{year}{2014}).
\newblock \urlprefix\url{http://www.aanda.org/10.1051/0004-6361/201424193}.
\newblock \eprint{arXiv:1405.2938v1}.

\bibitem{Melzani2013}
\bibinfo{author}{Melzani, M.} \emph{et~al.}
\newblock \bibinfo{title}{Apar-t: code, validation, and physical interpretation
  of particle-in-cell results}.
\newblock \emph{\bibinfo{journal}{Astronomy \& Astrophysics}}
  \textbf{\bibinfo{volume}{558}}, \bibinfo{pages}{A133} (\bibinfo{year}{2013}).

\bibitem{DelSanto2013}
\bibinfo{author}{Del~Santo, M.}, \bibinfo{author}{Malzac, J.},
  \bibinfo{author}{Belmont, R.}, \bibinfo{author}{Bouchet, L.} \&
  \bibinfo{author}{De~Cesare, G.}
\newblock \bibinfo{title}{The magnetic field in the x-ray corona of cygnus
  x-1}.
\newblock \emph{\bibinfo{journal}{Monthly Notices of the Royal Astronomical
  Society}} \textbf{\bibinfo{volume}{430}}, \bibinfo{pages}{209--220}
  (\bibinfo{year}{2013}).

\bibitem{Reis2013}
\bibinfo{author}{Reis, R.} \& \bibinfo{author}{Miller, J.}
\newblock \bibinfo{title}{On the size and location of the x-ray emitting
  coronae around black holes}.
\newblock \emph{\bibinfo{journal}{The Astrophysical Journal Letters}}
  \textbf{\bibinfo{volume}{769}}, \bibinfo{pages}{L7} (\bibinfo{year}{2013}).

\bibitem{Mckinney2011}
\bibinfo{author}{McKinney, J.~C.} \& \bibinfo{author}{Uzdensky, D.~A.}
\newblock \bibinfo{title}{A reconnection switch to trigger gamma-ray burst jet
  dissipation}.
\newblock \emph{\bibinfo{journal}{Monthly Notices of the Royal Astronomical
  Society}} \textbf{\bibinfo{volume}{419}}, \bibinfo{pages}{573--607}
  (\bibinfo{year}{2011}).

\bibitem{Arber2015}
\bibinfo{author}{Arber, T.} \emph{et~al.}
\newblock \bibinfo{title}{Contemporary particle-in-cell approach to
  laser-plasma modelling}.
\newblock \emph{\bibinfo{journal}{Plasma Physics and Controlled Fusion}}
  \textbf{\bibinfo{volume}{57}}, \bibinfo{pages}{113001}
  (\bibinfo{year}{2015}).

\bibitem{Sato2018}
\bibinfo{author}{Sato, T.} \emph{et~al.}
\newblock \bibinfo{title}{Features of particle and heavy ion transport code
  system (phits) version 3.02}.
\newblock \emph{\bibinfo{journal}{Journal of Nuclear Science and Technology}}
  \textbf{\bibinfo{volume}{55}}, \bibinfo{pages}{684--690}
  (\bibinfo{year}{2018}).

\bibitem{Chubar1998}
\bibinfo{author}{Chubar, O.}, \bibinfo{author}{Elleaume, P.} \&
  \bibinfo{author}{Chavanne, J.}
\newblock \bibinfo{title}{A three-dimensional magnetostatics computer code for
  insertion devices}.
\newblock \emph{\bibinfo{journal}{Journal of synchrotron radiation}}
  \textbf{\bibinfo{volume}{5}}, \bibinfo{pages}{481--484}
  (\bibinfo{year}{1998}).

\end{thebibliography}

\begin{acknowledgments}
We thank the technical staff of ILE (GEKKO XII, LFEX) for their support during experiments. This work was supported by fundings from ILE(2017A1-KORNEEV), MEXT, JSPS by Grant-in-Aid for JSPS Research Fellow (18J11119, 18J11354), Grants-in-Aid, KAKENHI (Grant No. 15KK0163, 15K21767, 16K13918, 16H02245) and Bilateral Program for Supporting International Joint Research, MEPhI by Academic Excellence Project (Contract No. 02.a03.21.0005-27.08.2013). This work was also partially supported by PRESTO (JPMJPR15PD) commissioned by JST. We appreciate for valuable discussions with T. Sano on MHD phenomena occurred in an accretion disk.
\end{acknowledgments}

\section{Author Contributions}
K. F. F. L. and Y. Ab. equally contributed to the experiment and analysis.
The paper was first drafted by K. F. F. L. and S. F., and then revised by specific comments from P. K., J. J. S., A. Y. and H. M.
S. F. and P. K. are the principal-co-investigators who proposed and organized this study.
The experiment was carried out with the help of S. S., S. L., K. M., H. M., Y. O., C. L., K. O., D. G. and E. M. under the supervisions by Y. Ar., A. Y., T. O., M. N., J. J. S., E. d'H., P. K. and S. F.
K. F. F. L. performed computer simulation and theoretical analysis under the supervision by A. M., Y. S. and P. K.
All authors contributed to the discussion of the results.

\section{Author Information}
The authors declare no competing interests.
Correspondence and requests for materials should be addressed to K. F. F. L. (k-law@ile.osaka-u.ac.jp) or S. F. (sfujioka@ile.osaka-u.ac.jp).

\section{Figures in Extended Data}

\end{document}